\documentclass[a4paper,11pt]{article}
\pdfoutput=1 

\usepackage{jheppub} 

\usepackage[T1]{fontenc} 

\centerline{\title{\boldmath One-particle reducible contribution to the one-loop scalar propagator  in a constant field}}

\author[a]{James P. Edwards}
\author[a]{and Christian Schubert}

\affiliation[a]{
Instituto de F\'{\i}sica y Matem\'aticas,
\\
Universidad Michoacana de San Nicol\'as de Hidalgo,\\
Edificio C-3, Apdo. Postal 2-82,\\
C.P. 58040, Morelia, Michoac\'an, M\'exico\\
}

\emailAdd{jedwards@ifm.umich.mx}
\emailAdd{schubert@ifm.umich.mx}

\abstract{Recently, Gies and Karbstein showed that the two-loop Euler-Heisenberg Lagrangian receives a finite one-particle reducible contribution
in addition to the well-known one-particle irreducible one. Here, we demonstrate that a similar contribution exists for the propagator in a constant field already at the
one-loop level, and we calculate this contribution for the scalar QED case. We also present an independent derivation of the Gies-Karbstein result
using the worldline formalism, treating the scalar and spinor QED cases in a unified manner. 
}

\begin{document} 
\maketitle
\flushbottom

\def\green{\color{green}}
\def\blue{\color{blue}}
\def\red{\color{red}}
\def\black{\color{black}}
%

\def\veps{\varepsilon}
\newcommand{\Zz}{\mathcal{Z}}
\newcommand{\Zzp}{\mathcal{Z}^{\prime}}
\newcommand{\detZ}{\textrm{det}^{-\frac{1}{2}}\left[\frac{\sin\Zz}{\mathcal{\Zz}}\right]}
\newcommand{\detZp}{\textrm{det}^{-\frac{1}{2}}\left[\frac{\sin \Zzp}{\Zzp}\right]}
\newcommand{\detZs}{\textrm{det}^{-\frac{1}{2}}\left[\frac{\tan\Zz}{\mathcal{\Zz}}\right]}
\newcommand{\detZps}{\textrm{det}^{-\frac{1}{2}}\left[\frac{\tan\Zzp}{\Zzp}\right]}
\newcommand{\pdetZ}{\textrm{det}^{-\frac{1}{2}}\left[\cos \Zz \right]}
\newcommand{\pdetZp}{\textrm{det}^{-\frac{1}{2}}\left[\cos(\Zzp)\right]}

\newcommand{\tZz}{\frac{\tan\Zz}{\Zz}}
\newcommand{\tZzp}{\frac{\tan\Zzp}{\Zzp}}

\newcommand{\xm}{x_{-}}
\newcommand{\xp}{x_{+}}
\newcommand{\yp}{y_{+}}
\newcommand{\ym}{y_{-}}

\newcommand{\delC}{\underset{\smile}{\Delta}}
\newcommand{\ddelC}{{^{\bullet}\!\delC}}
\newcommand{\delCd}{{\delC\!^{\bullet}}}
\newcommand{\ddelCd}{{^{\bullet}\!\delC\!^{\bullet}}}
\newcommand{\odelC}{{^{\circ}\!\delC}}
\newcommand{\delCo}{{\delC\!^{\circ}}}
\newcommand{\odelCo}{{^{\circ}\!\delC\!^{\circ}}}
\newcommand{\odelCd}{{^{\circ}\!\delC\!^{\bullet}}}
\newcommand{\ddelCo}{{^{\bullet}\!\delC\!^{\circ}}}

\newcommand{\gb}{{\mathcal{G}_{B}}}
\newcommand{\gbd}{{\dot{\mathcal{G}}_{B}}}
\newcommand{\gbdm}{\dot{\mathcal{G}}_{B \mu\nu}}
\newcommand{\gf}{{\mathcal{G}_{F}}}
\newcommand{\gfd}{{\dot{\mathcal{G}}_{F}}}
\newcommand{\gfdm}{\dot{\mathcal{G}}_{F \mu\nu}}

\def\cosech{\rm cosech}
\def\sech{\rm sech}
\def\coth{\rm coth}
\def\tanh{\rm tanh}
\def\tan{\rm tan}
\def\half{{1\over 2}}
\def\third{{1\over3}}
\def\fourth{{1\over4}}
\def\fifth{{1\over5}}
\def\sixth{{1\over6}}
\def\seventh{{1\over7}}
\def\eigth{{1\over8}}
\def\ninth{{1\over9}}
\def\tenth{{1\over10}}
\def\conj{{{\rm c.c.}}}
\def\bN{\mathop{\bf N}}
\def\R{{\rm I\!R}}
\def\Eins{{\mathchoice {\rm 1\mskip-4mu l} {\rm 1\mskip-4mu l}
{\rm 1\mskip-4.5mu l} {\rm 1\mskip-5mu l}}}
\def\Z{{\mathchoice {\hbox{$\sf\textstyle Z\kern-0.4em Z$}}
{\hbox{$\sf\textstyle Z\kern-0.4em Z$}}
{\hbox{$\sf\scriptstyle Z\kern-0.3em Z$}}
{\hbox{$\sf\scriptscriptstyle Z\kern-0.2em Z$}}}}
\def\abs#1{\left| #1\right|}
\def\com#1#2{
        \left[#1, #2\right]}
\def\contract{\makebox[1.2em][c]{
        \mbox{\rule{.6em}{.01truein}\rule{.01truein}{.6em}}}}
\def\ltap{\ \raisebox{-.4ex}{\rlap{$\sim$}} \raisebox{.4ex}{$<$}\ }
\def\gtap{\ \raisebox{-.4ex}{\rlap{$\sim$}} \raisebox{.4ex}{$>$}\ }
\def\mn{{\mu\nu}}
\def\rs{{\rho\sigma}}
\newcommand{\Det}{{\rm Det}}
\def\Tr{{\rm Tr}\,}
\def\tr{{\rm tr}\,}
\def\sumij{\sum_{i<j}}
\def\e{\,{\rm e}}
\def\br{{\bf r}}
\def\bp{{\bf p}}
\def\bq{{\bf q}}
\def\bx{{\bf x}}
\def\by{{\bf y}}
\def\brhat{{\bf \hat r}}
\def\bv{{\bf v}}
\def\ba{{\bf a}}
\def\bE{{\bf E}}
\def\bB{{\bf B}}
\def\bA{{\bf A}}
\def\b0{{\bf 0}}
\def\pa{\partial}
\def\dA{\partial^2}
\def\ddx{{d\over dx}}
\def\ddt{{d\over dt}}
\def\der#1#2{{d #1\over d#2}}
\def\lie{\hbox{\it \$}} 
\def\partder#1#2{\frac{\partial #1}{\partial #2}}
\def\secder#1#2#3{{\partial^2 #1\over\partial #2 \partial #3}}
%
\def\be{\begin{equation}}
\def\ee{\end{equation}\noindent}
\def\bear{\begin{eqnarray}}
\def\ear{\end{eqnarray}\noindent}
\def\bec{\blue\begin{equation}}
\def\eec{\end{equation}\black\noindent}
\def\bearc{\blue\begin{eqnarray}}
\def\earc{\end{eqnarray}\black\noindent}
\def\benn{\begin{enumerate}}
\def\enn{\end{enumerate}}
\def\veject{\vfill\eject}
\def\ven{\vfill\eject\noindent}
%
\def\eq#1{{eq. (\ref{#1})}}
\def\eqs#1#2{{eqs. (\ref{#1}) -- (\ref{#2})}}
%
\def\inv#1{\frac{1}{#1}}
\def\sumninf{\sum_{n=0}^{\infty}}
%
\def\totint{\int_{-\infty}^{\infty}}
\def\posint{\int_0^{\infty}}
\def\negint{\int_{-\infty}^0}
\def\pint{{\dps\int}{dp_i\over {(2\pi)}^d}}
\def\intdp3{\int\frac{d^3p}{(2\pi)^3}}
\def\intdp4{\int\frac{d^4p}{(2\pi)^4}}
\def\scalprop#1{\frac{-i}{#1^2+m^2-i\epsilon}}
%
\newcommand{\GeV}{\mbox{GeV}}
\def\FFdual{F\cdot\tilde F}
\def\bra#1{\langle #1 |}
\def\ket#1{| #1 \rangle}
\def\braket#1#2{\langle {#1} \mid {#2} \rangle}
\def\vev#1{\langle #1 \rangle}
\def\matel#1#2#3{\langle #1\mid #2\mid #3 \rangle}
\def\rightvac{\mid0\rangle}
\def\leftvac{\langle0\mid}
\def\ihbar{{i\over\hbar}}
\def\lagr{{\cal L}}
\def\sigmabar{{\bar\sigma}}
\def\ge{\hbox{$\gamma_1$}}
\def\gz{\hbox{$\gamma_2$}}
\def\gd{\hbox{$\gamma_3$}}
\def\go{\hbox{$\gamma_1$}}
\def\gt{\hbox{\$\gamma_2$}}
\def\gth{\hbox{$\gamma_3$}} 
\def\gf{\hbox{$\gamma_5\;$}}
\def\slash#1{#1\!\!\!\raise.15ex\hbox {/}}
\newcommand{\slD}{\,\raise.15ex\hbox{$/$}\kern-.27em\hbox{$\!\!\!D$}}
\newcommand{\slpartial}{\raise.15ex\hbox{$/$}\kern-.57em\hbox{$\partial$}}
\newcommand{\PP}{\cal P}
\newcommand{\G}{{\cal G}}
\newcommand{\nc}{\newcommand}
\nc{\spa}[3]{\left\langle#1\,#3\right\rangle}
\nc{\spb}[3]{\left[#1\,#3\right]}
\nc{\ksl}{\not{\hbox{\kern-2.3pt $k$}}}
\nc{\hf}{\textstyle{1\over2}}
\nc{\pol}{\varepsilon}
\nc{\tq}{{\tilde q}}
\nc{\esl}{\not{\hbox{\kern-2.3pt $\pol$}}}
\newcommand{\cL}{\cal L}
\newcommand{\D}{\cal D}
\newcommand{\Dhalf}{{D\over 2}}
\def\eps{\epsilon}
\def\epshalf{{\epsilon\over 2}}
\def\lag{( -\partial^2 + V)}
\def\freeexp{{\rm e}^{-\int_0^Td\tau {1\over 4}\dot x^2}}
\def\kinb{{1\over 4}\dot x^2}
\def\kinf{{1\over 2}\psi\dot\psi}
\def\expk{{\rm exp}\biggl[\,\sum_{i<j=1}^4 G_{Bij}k_i\cdot k_j\biggr]}
\def\expp{{\rm exp}\biggl[\,\sum_{i<j=1}^4 G_{Bij}p_i\cdot p_j\biggr]}
\def\expshort{{\e}^{\half G_{Bij}k_i\cdot k_j}}
\def\expabb{{\e}^{(\cdot )}}
\def\epseps#1#2{\varepsilon_{#1}\cdot \varepsilon_{#2}}
\def\epsk#1#2{\varepsilon_{#1}\cdot k_{#2}}
\def\kk#1#2{k_{#1}\cdot k_{#2}}
\def\G#1#2{G_{B#1#2}}
\def\Gp#1#2{{\dot G_{B#1#2}}}
\def\GF#1#2{G_{F#1#2}}
\def\Dab{{(x_a-x_b)}}
\def\Dsq{{({(x_a-x_b)}^2)}}
\def\PITD{{(4\pi T)}^{-{D\over 2}}}
\def\4piTD{{(4\pi T)}^{-{D\over 2}}}
\def\4piT4{{(4\pi T)}^{-2}}
\def\TintmD{{\dps\int_{0}^{\infty}}{dT\over T}\,e^{-m^2T}
    {(4\pi T)}^{-{D\over 2}}}
\def\Tintm4{{\dps\int_{0}^{\infty}}{dT\over T}\,e^{-m^2T}
    {(4\pi T)}^{-2}}
\def\Tintm{{\dps\int_{0}^{\infty}}{dT\over T}\,e^{-m^2T}}
\def\Tint{{\dps\int_{0}^{\infty}}{dT\over T}}
\def\np{n_{+}}
\def\nm{n_{-}}
\def\Np{N_{+}}
\def\Nm{N_{-}}
\newcommand{\slG}{{{\dot G}\!\!\!\! \raise.15ex\hbox {/}}}
\newcommand{\Gd}{{\dot G}}
\newcommand{\Gund}{{\underline{\dot G}}}
\newcommand{\Gdd}{{\ddot G}}
\def\GBd12{{\dot G}_{B12}}
\def\Dx{\dps\int{\cal D}x}
\def\Dy{\dps\int{\cal D}y}
\def\Dpsi{\dps\int{\cal D}\psi}
\def\dint#1{\int\!\!\!\!\!\int\limits_{\!\!#1}}
\def\ddtau{{d\over d\tau}}
\def\ie{\hbox{$\textstyle{\int_1}$}}
\def\iz{\hbox{$\textstyle{\int_2}$}}
\def\id{\hbox{$\textstyle{\int_3}$}}
\def\ldop{\hbox{$\lbrace\mskip -4.5mu\mid$}}
\def\rdop{\hbox{$\mid\mskip -4.3mu\rbrace$}}
%
\newcommand{\1}{{\'\i}}
\newcommand{\no}{\noindent}
\def\non{\nonumber}
\def\dps{\displaystyle}
\def\sy{\scriptscriptstyle}
\def\sy{\scriptscriptstyle}

\section{Introduction}
\label{sec:intro}

The Euler-Heisenberg Lagrangian (`EHL'), one of the first serious calculations
in quantum electrodynamics \cite{eulhei}, describes the
one-loop amplitude involving a spinor loop
interacting non-perturbatively with a constant background
electromagnetic field. It is also the first, and prototypical, example
of an effective Lagrangian in field theory. Euler and 
Heisenberg found for it the following well-known integral representation: 

\bear
{\cal L}_{\rm spin}^{(1)} (a,b)
&=&
-
{1\over 8\pi^2}
\int_0^{\infty}{dT\over T}
\,\e^{-m^2T}
{e^2ab\over \tanh(eaT)\tan(ebT)} \, .
\non\\
\label{L1spin}
\ear
Here $T$ denotes the (Euclidean) proper-time of the
loop fermion, and $a,b$ are related to the two
invariants of the Maxwell field by
$a^2-b^2=B^2-E^2, ab = {\bf E}\cdot{\bf B}$.
The superscript `(1)' stands for `one-loop'.

The EHL contains the information on nonlinear QED effects \cite{giesbook} such 
as photon--photon scattering \cite{eulhei,itzzub-book,97,rebtur,elmayo},
photon dispersion \cite{toll,adler71}, and photon splitting
\cite{biabia,adler71,bamish,adlsch}.
Combined with spinor helicity methods, it allows one to explicitly
compute the low-energy limit of the $N$-photon amplitudes for arbitrary $N$ in the helicity decomposition \cite{56}. 
Its imaginary part holds the information on Sauter-Schwinger pair production \cite{schwinger51}.
See \cite{dunneeulhei} for a review of the uses of the EHL.

For scalar QED, an analogous result was obtained by Weisskopf and
Schwinger \cite{weisskopf,schwinger51}:

\bear
{\cal L}_{\rm scal}^{(1)} (a,b)
&=&
{1\over 16\pi^2}
\int_0^{\infty}{dT\over T}
\,\e^{-m^2T}
{e^2ab\over \sinh(eaT)\sin(ebT)} \, .
\non\\
\label{L1scal}
\ear
However, we will call this Lagrangian ``scalar EHL'' for simplicity. 

The Lagrangians \eqref{L1spin},\eqref{L1scal} require renormalization, but we will
not bother here to include the corresponding counterterms, since they will
not play a role as far as the present paper is concerned. 

The first radiative corrections to these Lagrangians,
describing the effect of an additional photon exchange in the loop, were
obtained in the seventies by Ritus \cite{ritusspin,ritusscal,ginzburg}. 
Using the exact propagators in a constant field \cite{fock,schwinger51},  Ritus obtained the 
corresponding two-loop effective Lagrangians ${\cal L}_{\rm scal,spin}^{(2)}$
in terms of certain two-parameter integrals.
Similar two-parameter integral
representations for ${\cal L}_{\rm scal,spin}^{(2)}$ were obtained
later by other authors, using either 
proper-time \cite{ditreuqed,18}
or dimensional regularisation \cite{frss,korsch}.
Although these integrals are intractable analytically, closed-form expressions
have been obtained for their weak-field expansions for the purely electric
or magnetic cases \cite{37,66}.
In 1+1 dimensions, the (spinor) EHL has also been calculated at the three-loop level, leading to four-parameter
integrals \cite{81,85,ip}. 

\begin{figure}[htbp]
\begin{center}
 \includegraphics[width=0.16\textwidth]{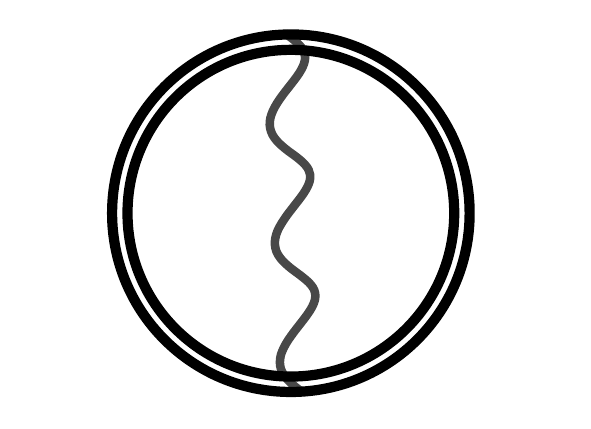}
\caption{{\bf One-particle irreducible contribution to the two-loop EHL.}}
\label{fig-EHL1PI}
\end{center}
\end{figure}

In all these calculations it was assumed that the only diagram contributing to the EHL at the two-loop
level is the one particle irreducible (`1PI') one shown in Fig. \ref{fig-EHL1PI} (the double line as usual denotes the 
electron propagator in a constant field).

\vspace{20pt}

\begin{figure}[htbp]
\begin{center}
 \includegraphics[width=0.35\textwidth]{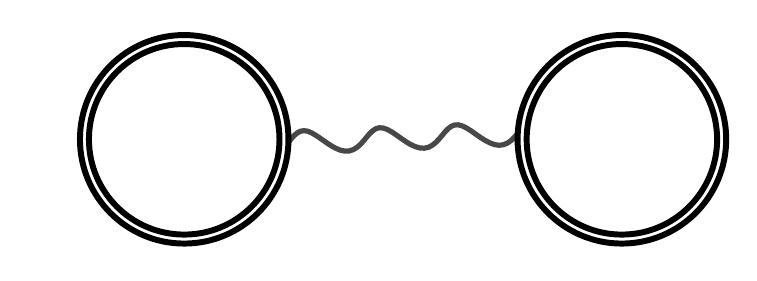}
\caption{{\bf One-particle reducible contribution to the two-loop EHL.}}
\label{fig-EHL1PR}
\end{center}
\end{figure}

\noindent
At the same loop order, there is also the one-particle reducible (`1PR') diagram shown in Fig. \ref{fig-EHL1PR}.

\noindent
However, since the one-photon amplitude in a constant field formally vanishes on account of momentum conservation,
this diagram was generally believed not to contribute. 

Recently, Gies and Karbstein \cite{giekar} made the stunning discovery that actually this diagram {\sl does} give a finite contribution,
if one takes into account the divergence of the connecting photon propagator in the zero-momentum limit. A careful analysis of that
limit led them to the simple formula

\bear
{\cal L}^{(2)\rm 1PR}
&=& \partder{{\cal L}^{(1)}}{F^\mn} \partder{{\cal L}^{(1)}}{F_\mn} \, .
\label{gieskarb}
\ear
Thus essentially all previous applications of the two-loop EHL now have to be reanalyzed to take the effects of this term into account.

\begin{figure}[htbp]
\begin{center}
 \includegraphics[width=0.35\textwidth]{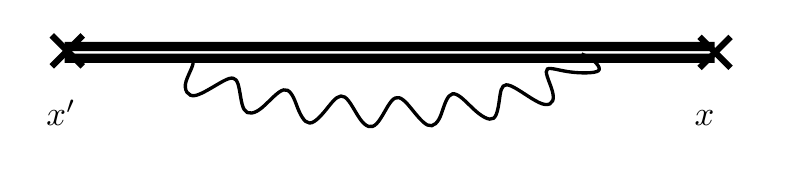}
\caption{{\bf One-loop propagator in a constant field.}}
\label{fig-SE1PI}
\end{center}
\end{figure}

The purpose of the present paper is to point out that a similar 1PR addendum exists also for the
QED scalar or electron propagators, here already at the one-loop level: usually in a constant field one considers only
the 1PI diagram shown in Fig. \ref{fig-SE1PI}, 

\hspace{20pt}

\noindent
but there is also a finite contribution from the 1PR one shown in Fig. \ref{fig-SE1PR} below. 
For the scalar QED case, we will explicitly obtain the contribution of this diagram, and obtain a generalization of
\eqref{gieskarb}. 

As a preparation for that calculation, we will first rederive the Gies-Karbstein result in a way that
allows us to combine the scalar and spinor QED cases, using the worldline formalism \cite{strassler,5,15,shaisultanov,18,40,41} (only the spinor QED case was considered in \cite{giekar}).

\section{Worldline rederivation of the Gies-Karbstein addendum to the EHL}

\subsection{Scalar QED}

The worldline path integral representation of the one-loop one-photon scalar QED amplitude
in a constant field $F_\mn$ is \cite{18,40,41}

\bear
\Gamma_{\rm scal}^{(1)}[k,\varepsilon;F] = 
-ie\Tintm \int_{x(T) = x(0)} \!\!\!\!\!\!\!\!\!\!Dx \int_0^Td\tau \varepsilon\cdot \dot x \e^{ik\cdot x(\tau)}\, \e^{-\int_0^T d\tau \bigl( \frac{\dot x^2}{4} + ie\dot x\cdot A \bigr)} \, .
\nonumber\\
\label{Gammascal}
\ear
Here the momentum $k$ is ingoing. Separating off the loop center of mass $x_0$,

\bear
x(\tau) = x_0 + q(\tau)
\ear
and choosing Fock-Schwinger gauge centered at $x_0$, one has \cite{5}

\bear
\Gamma_{\rm scal}^{(1)}[k,\varepsilon;F] = -ie
\int d^Dx_0 \e^{ik\cdot x_0} \Tintm \int_{q(0)=0}^{q(T)=0}
\hspace{-13pt} Dq \int_0^Td\tau \varepsilon\cdot \dot q \e^{ik\cdot q(\tau)}
\, \e^{-\int_0^T d\tau \bigl( \frac{\dot q^2}{4} + \half  ie q\cdot F \cdot \dot q \bigr)}
\, .
\nonumber\\
\ear
The $x_0$ integration produces the delta function for momentum conservation. 
Performing the gaussian path integration along the lines of \cite{strassler,18,40} gives 

\begin{eqnarray}
\Gamma_{\rm scal}^{(1)}
[k,\varepsilon;F]
&=&
-e
{(2\pi )}^D\delta^D (k)
\int_{0}^{\infty}}{dT\over T}
{(4\pi T)^{-{D\over 2}}
e^{-m^2T}
{\rm det}^{-{1\over 2}}
\biggl[{{\rm sin}{\cal Z}\over {\cal Z}}\biggr]
\nonumber\\ 
&& \times
\int_0^T  d\tau 
\varepsilon\cdot\dot{\cal G}_{B}\cdot k
\,{\rm exp}\Bigl[\half k\cdot {\cal G}_{B}\cdot  k\Bigr]
\nonumber\\
\label{1photonscalT}
\end{eqnarray}
where ${\cal Z}_{\mu\nu} \equiv eF_{\mu\nu}T$, and ${\cal G}_B$, $\dot{\cal G}_B$ 
are the coincidence limits of the
constant field worldline Green's function and its first derivative:

\begin{eqnarray}
{\cal G}_B &\equiv & {\cal G}_B(\tau,\tau) =
{T\over 2{\cal Z}^2}
\Bigl({\cal Z}\cdot \cot {\cal Z}- 1\Bigr)\, , \nonumber\\
\dot {\cal G}_B &\equiv& \dot {\cal G}_B(\tau,\tau)  = i{\rm cot}{\cal Z}
-{i\over {\cal Z}} = i\frac{2{\cal Z}}{T}{\cal G}_B \, .
\nonumber\\
\label{coin}
\end{eqnarray}
Since these coincidence limits are independent of $\tau$, we can simplify (\ref{1photonscalT}) to 

\begin{eqnarray}
\Gamma_{\rm scal}^{(1)}[k,\varepsilon;F]
&=&
-e
{(2\pi )}^D\delta^D (k)
\int_{0}^{\infty}}dT
{(4\pi T)^{-{D\over 2}}
e^{-m^2T}
{\rm det}^{-{1\over 2}}
\biggl[{{\rm sin}{\cal Z}\over {\cal Z}}\biggr]
\nonumber\\ 
&& \times
\varepsilon\cdot\dot{\cal G}_{B}\cdot k
\,{\rm exp}\Bigl[\half k\cdot {\cal G}_{B}\cdot  k\Bigr] \, .
\nonumber\\
\label{1photonscal}
\end{eqnarray}
From \eqref{1photonscal}, we can now simply construct the reducible diagram \ref{fig-EHL1PR} by taking two copies of it, and connecting them with
a photon propagator in Feynman gauge. This gives

\bear
\Gamma_{\rm scal}^{(2)\rm 1PR}
&=& \int \frac{d^Dk}{(2 \pi)^{D}k^2} 
 \Gamma_{\rm scal}^{(1)}[k,\varepsilon;F]\Gamma_{\rm scal}^{(1)}[k',\varepsilon';F]\Bigl\vert_{k'\to -k,\pol^{\mu}\pol'^{\nu}\to \eta^\mn} 
\nonumber\\
&=& 
e^2
 \int \frac{d^Dk}{(2\pi)^D k^2} 
\int d^Dx_0 \e^{ik\cdot x_0}
\int_{0}^{\infty}}dT
{(4\pi T)^{-{D\over 2}}
e^{-m^2T}
{\rm det}^{-{1\over 2}}
\biggl[{{\rm sin}{\cal Z}\over {\cal Z}}\biggr]
\nonumber\\&&\times
\int d^Dx'_0 \e^{-ik\cdot x'_0}
\int_{0}^{\infty}}dT'
{(4\pi T')^{-{D\over 2}}
e^{-m^2T'}
{\rm det}^{-{1\over 2}}
\biggl[{{\rm sin}{\cal Z'}\over {\cal Z'}}\biggr]
\nonumber\\&&\times
k\cdot\dot{\cal G}_{B}\cdot\dot{\cal G}'_{B}\cdot k
\,{\rm exp}\Bigl[\half k\cdot ({\cal G}_{B}+ {\cal G}'_{B})\cdot  k\Bigr] \, .
\label{sewing}
\ear
Here we have used that $\dot{\cal G}_{B}(\tau,\tau)$ is an odd function of the antisymmetric matrix $F$, and therefore antisymmetric too. 

Now we change variables from $x_0,x_0'$ to $x_+ \equiv \half(x_0+x'_0)$, $x_- \equiv x_0 -x'_0$. 
The center of mass $x_+$ remains fixed, and will be the argument of the two-loop effective Lagrangian. 
Integrating out $x_-$, we get

\bear
{\cal L}_{\rm scal}^{(2)\rm 1PR} (x\equiv x_+) &=& e^2 
\int_0^{\infty}dT
(4\pi T)^{-{D\over 2}}
e^{-m^2T}
{\rm det}^{-{1\over 2}}
\biggl[{{\rm sin}{\cal Z}\over {\cal Z}}\biggr]
\nonumber\\&&\times
\int_0^{\infty}dT'
(4\pi T')^{-{D\over 2}}
e^{-m^2T'}
{\rm det}^{-{1\over 2}}
\biggl[{{\rm sin}{\cal Z'}\over {\cal Z'}}\biggr]
\nonumber\\&&\times
 \int d^Dk \, \delta^D(k)\frac{ k\cdot\dot{\cal G}_{B}\cdot\dot{\cal G}'_{B}\cdot k}{k^2}
\,{\rm exp}\Bigl[\half k\cdot ({\cal G}_{B}+ {\cal G}'_{B})\cdot  k\Bigr] \, .
\nonumber\\
\ear
The crucial observation of \cite{giekar} was that, although the presence of the delta function would seem
to kill the one-photon amplitude \eqref{1photonscal}, after the sewing this is not the case anymore; terms with only two factors of momentum
in the numerator will survive the momentum integration, and give a finite result. 
Thus, in the presence of the delta function, the exponential factor in the last line can be replaced by unity. 
Lorentz invariance can then be invoked to set

\bear
 \int d^Dk \, \delta^D(k) \frac{k^{\mu}k^{\nu}}{k^2} = \frac{\eta^\mn}{D} \, .
 \label{intk}
\ear
Thus one has finally
 
 \bear
{\cal L}_{\rm scal}^{(2)\rm 1PR} &=& e^2 
\int_0^{\infty}dT
(4\pi T)^{-{D\over 2}}
e^{-m^2T}
{\rm det}^{-{1\over 2}}
\biggl[{{\rm sin}{\cal Z}\over {\cal Z}}\biggr]
\nonumber\\&&\times
\int_0^{\infty}dT'
(4\pi T')^{-{D\over 2}}
e^{-m^2T'}
{\rm det}^{-{1\over 2}}
\biggl[{{\rm sin}{\cal Z'}\over {\cal Z'}}\biggr]
\nonumber\\&&\times
\frac{1}{D}{\rm tr}\bigl(\dot{\cal G}_{B}\cdot\dot{\cal G}'_{B} \bigr) \, .
\ear
Now, we use the fact that the scalar EHL \eqref{L1scal} can also be rewritten more compactly as
\cite{5}

 \bear
{\cal L}_{\rm scal}^{(1)}(F) &=&  
\int_0^{\infty}\frac {dT}{T}
(4\pi T)^{-{D\over 2}}
e^{-m^2T}
{\rm det}^{-{1\over 2}}
\biggl[{{\rm sin}{\cal Z}\over {\cal Z}}\biggr] \, .
\nonumber\\
\label{L1scalF}
\ear
Using the identity $\ln \det = \tr \ln $ as well as

\bear
-i\dot{\cal G}_{B} = \cot {\cal Z} - \frac{1}{\cal Z} = \frac{d}{d{\cal Z}} \ln \frac{\sin {\cal Z} }{\cal Z} 
\label{idtrigscal}
\ear
and setting $D=4$, it is then easy to verify the Gies-Karbstein equation \eqref{gieskarb}.

\subsection{Spinor QED}

The generalization of the one-photon amplitude \eqref{1photonscal} to the spinor-loop case
in the worldline formalism involves, apart from the global normalization, only a change
of the determinant factor, and an application of the Bern-Kosower rules \cite{berkos-prl,berkos-npb} 
to replace $\dot{\cal G}_{B}$ by $\dot{\cal G}_{B}-{\cal G}_{F}$, where 
${\cal G}_F = -i\tan {\cal Z}$:

\begin{eqnarray}
\Gamma_{\rm spin}^{(1)}
[k,\varepsilon;F]
&=&
2e
\int d^Dx_0 \e^{ik\cdot x_0}
\int_{0}^{\infty}}{dT\over T}
{(4\pi T)^{-{D\over 2}}
e^{-m^2T}
{\rm det}^{-{1\over 2}}
\biggl[{{\rm tan}{\cal Z}\over {\cal Z}}\biggr]
\nonumber\\ 
&& \times
\int_0^T  d\tau 
\varepsilon\cdot(\dot{\cal G}_{B}-{\cal G}_F)\cdot k
\,{\rm exp}\Bigl[\half k\cdot {\cal G}_{B}\cdot  k\Bigr] \, .
\nonumber\\
\label{1photonspin}
\end{eqnarray}
The procedure is the same as in the scalar case {\it mutatis mutandis}: 
the spinor QED EHL should be rewritten as \cite{5}

 \bear
{\cal L}_{\rm spin}^{(1)}(F) &=&  
-2\int_0^{\infty}\frac {dT}{T}
(4\pi T)^{-{D\over 2}}
e^{-m^2T}
{\rm det}^{-{1\over 2}}
\biggl[{{\rm tan}{\cal Z}\over {\cal Z}}\biggr]
\nonumber\\
\label{L1spinF}
\ear
and  instead of the identity \eqref{idtrigscal} one has to use
 
\bear
-i(\dot{\cal G}_{B}-{\cal G}_{F}) 
 = \cot {\cal Z} - \frac{1}{\cal Z}
 +\tan {\cal Z} = \frac{d}{d{\cal Z}} \ln \frac{\tan {\cal Z} }{\cal Z}  \, .
 \label{idtrigspin}
\ear
One then finds the same identity as in the scalar case, eq. \eqref{gieskarb}.

\section{An analogous addendum for the scalar propagator}

We will now generalize the Gies-Karbstein addendum to the case where, instead of 
two closed loops, we have one closed loop and an open line, i.e. to the 
self-energy diagram shown in Fig. \ref{fig-SE1PR}. Here we will restrict ourselves to the scalar QED case. 

\begin{figure}[htbp]
\begin{center}
 \includegraphics[width=0.35\textwidth]{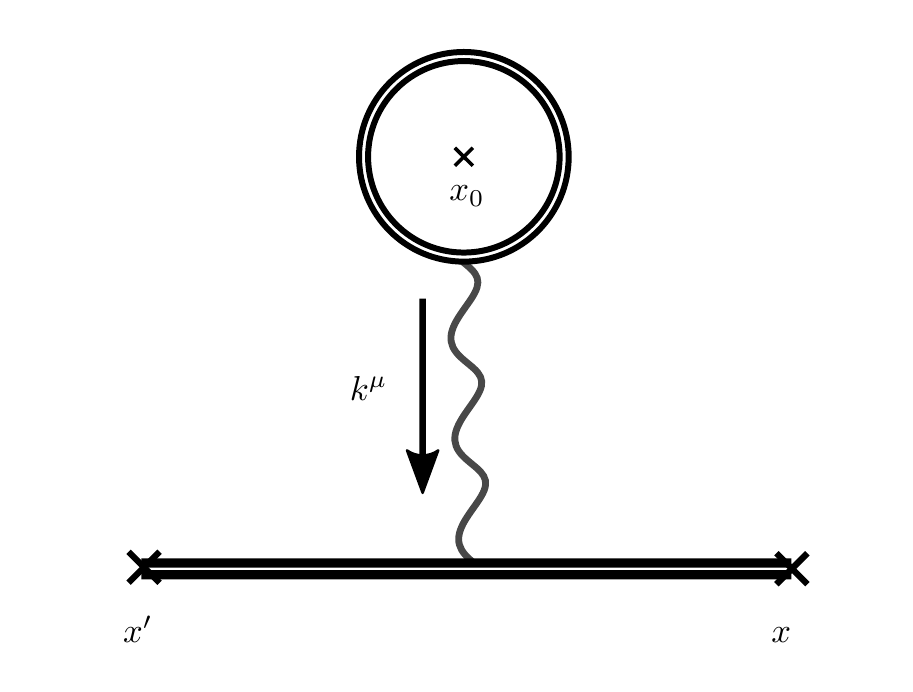}
\caption{{\bf One-particle reducible contribution to the one-loop scalar propagator in a constant field.}}
\label{fig-SE1PR}
\end{center}
\end{figure}

\subsection{The propagator in configuration space}

Let us start in $x$-space. 
In addition to the one-photon amplitude in the field, we now also need the scalar propagator $D^{xx'}[k,\varepsilon;F]$ between points $x$ and $x'$, in the constant field background
and with one photon attached. Similarly to \eqref{sewing}, we can then construct
the 1PR self-energy diagram $D^{xx'(1)1PR}$ by sewing,

\bear
D_{\rm scal}^{xx'(1)1PR}
&=& \int \frac{d^Dk}{(2\pi)^{D}k^2} 
 \Gamma_{\rm scal}^{(1)}[k',\varepsilon';F]D_{\rm scal}^{xx'}[k,\varepsilon;F]
 \Bigl\vert_{k'\to -k,\,\pol^{\mu}\pol'^{\nu}\to \eta^\mn}  \, .
\nonumber\\
\label{sewingGammaD}
\ear
In the worldline formalism, 
the amplitude $D^{xx'}[k,\varepsilon;F]$ can be obtained from the open-line master formula of \cite{110}.
Specializing eq. (3.7) there to the one-photon case, one gets

\bear
D^{xx'}[k,\varepsilon;F] 
	 &=& (-ie) \e^{i k \cdot x^{\prime}} \int_{0}^{\infty}dT e^{-m^{2}T}(4\pi T)^{-\frac{D}{2}} \detZ e^{-\frac{1}{4T} x_{-} \Zz \cdot \cot \Zz \cdot x_{-}} \nonumber \\
	&& \hspace{-70pt}\times \varepsilon_{\mu}\int_{0}^{T}d\tau
	\left(\frac{\xm}{T} - 2i \ddelC(\tau, \tau)\cdot k - \frac{2ie}{T}\xm \cdot F \cdot \odelCd(\tau) \right)^{\mu} 
	e^{i  \frac{\tau}{T}  k \cdot \xm+ k \cdot \delC(\tau, \tau) \cdot k + \frac{2e}{T}\xm \cdot F \cdot \odelC(\tau) \cdot k} \, .
	\nonumber\\
	\label{Dwl}
\ear
Here $x_-\equiv x-x'$, and instead of the worldline Green's function $\mathcal{G}_{B}(\tau, \tau^{\prime})$,
which fulfills the ``string-inspired condition'' (`SI')
$\int_{0}^{T}d\tau (d\tau') {\cal G}_{B}(\tau, \tau^{\prime}) = 0$, we encounter the corresponding
Green's function $\delC(\tau, \tau^{\prime})$ obeying Dirichlet boundary conditions (`DBC')
$\delC(0, \tau^{\prime})=\delC(\tau, 0) = 0$.
Since those boundary conditions break the translation invariance in $\tau$,
this Green's function is not a function of the difference of the arguments, 
so that one has to distinguish between its two partial derivatives.
A convenient notation \cite{basvanbook,110} is to 
denote differentiation with respect to $\tau$ ($\tau^{\prime}$) by a $\bullet$,
and the definite integral $\int_0^Td\tau$ ($\int_0^Td\tau'$) by a $\circ$ 
to the left (right) respectively. 
For the same reason, also the coincidence limits of $\delC(\tau, \tau^{\prime})$ and its
derivatives are not constants any more, so that the $\tau$-integral will have to be dealt
with.

From the sewing relation \eqref{sewingGammaD} together with \eqref{1photonscal} and \eqref{intk} it is clear
that only the terms linear in $k$ in the integrand of \eqref{Dwl} will survive the sewing (there is also a term independent of $k$,
but it can  be omitted, since it leads to a $k$-integral that vanishes by antisymmetry). Collecting those gives

\bear
D^{xx'}[k,\varepsilon;F]\Bigl\vert_{k} 
	 &=& (-ie) \int_{0}^{\infty}dT e^{-m^{2}T}(4\pi T)^{-\frac{D}{2}} \detZ e^{-\frac{1}{4T} x_{-} \Zz \cdot \cot \Zz \cdot x_{-}} \nonumber \\
	&& \hspace{-80pt}\times \int_{0}^{T}d\tau
	k\cdot
	\biggl\lbrack
	 \left(ix' + i  \frac{\tau}{T}\xm - \frac{2e}{T}\odelC^T(\tau) \cdot F \cdot \xm \right)
	 \left(\frac{\xm}{T} - \frac{2ie}{T}\xm \cdot F \cdot \odelCd(\tau)\right)
	- 2i \ddelC^T(\tau, \tau)
		 \biggr\rbrack
	\cdot\varepsilon\, .\nonumber\\
	\label{Dlin}
\ear
Combining this equation with \eqref{sewingGammaD}, \eqref{1photonscal} and \eqref{intk} we get

\bear
D_{\rm scal}^{xx'(1)1PR}
&=& -i\frac{e^2}{D}
\int_{0}^{\infty}}dT'
{(4\pi T')^{-{D\over 2}}
e^{-m^2T'}
{\rm det}^{-{1\over 2}}
\biggl[{{\rm sin}{\cal Z}'\over {\cal Z}'}\biggr]
\nonumber\\
&&\hspace{-25pt}\times
 \int_{0}^{\infty}dT e^{-m^{2}T}(4\pi T)^{-\frac{D}{2}} \detZ e^{-\frac{1}{4T} x_{-} \Zz \cdot \cot \Zz \cdot x_{-}} \nonumber\\
 &&\hspace{-25pt}\times
\tr \biggl\lbrace
 \dot{\cal G}'_{B} \cdot
\biggl\lbrack
	 \left(ix' + i  \frac{\tau}{T}\xm - \frac{2e}{T}\odelC^T(\tau) \cdot F \cdot \xm \right)
	 \left(\frac{\xm}{T} - \frac{2ie}{T}\xm \cdot F \cdot \odelCd(\tau)\right)
	- 2i \ddelC^T(\tau, \tau)
		 \biggr\rbrack
	\biggr\rbrace \, .
\nonumber\\ 
\label{sewed}
\ear
For the calculation of the $\tau$ integral, it will be convenient to rewrite the DBC Green's function $\delC$ in terms of the
SI one $\mathcal{G}_{B}$, using the relation \cite{110}

\begin{equation}
       2	\delC(\tau, \tau^{\prime}) = \mathcal{G}_{B}(\tau, \tau^{\prime}) - \mathcal{G}_{B}(\tau, 0) - \mathcal{G}_{B}(0, \tau^{\prime}) + \mathcal{G}_{B}(0,0)  \, .
\end{equation}
Using the collection of formulas given in \cite{110}, it is then easy to compute (note that all matrices appearing in the integrand are built from the same
$F_\mn$, and thus all commute with each other):

\begin{align}
	\int_{0}^{T}d\tau (\odelC)^{T} (\tau) &= \frac{T^{2}}{2}\mathcal{G}_{B}\, , \\
	\int_{0}^{T}d\tau \odelCd (\tau) &= 0 \, ,\\
	\int_{0}^{T}d\tau \odelCd (\tau) \tau &= -\frac{T^{2}}{2}\mathcal{G}_{B} \, ,  \\
	\int_{0}^{T}d\tau (\ddelC)^{T}(\tau,\tau) &= -\frac{T}{2}\dot{\mathcal{G}}_{B} \, ,\\
	\int_{0}^{T}d\tau \odelCd (\tau)  \cdot (\odelC)^{T} (\tau) 
	&= \frac{iT^{4}}{8\mathcal{Z}^{2}}\cdot \left[2i \dot{\mathcal{G}}_{B} +\left(\cot(\mathcal Z) - \mathcal{Z}\cdot \csc^{2} \mathcal{Z}\right)\right] \, .
\end{align}
Here it is understood that, on the right-hand side, $\mathcal{G}_{B}$ and $\dot{\mathcal{G}}_{B}$ are taken at their coincidence limits, given in
\eqref{coin}. Putting things together, and now also setting $D=4$, we have

\bear
D_{\rm scal}^{xx'(1)1PR}
&=&- i\frac{e^2}{4}
\int_{0}^{\infty}}dT'
{(4\pi T')^{-{D\over 2}}
e^{-m^2T'}
{\rm det}^{-{1\over 2}}
\biggl[{{\rm sin}{\cal Z'}\over {\cal Z'}}\biggr]
\nonumber\\
&&\times
 \int_{0}^{\infty}dT e^{-m^{2}T}(4\pi T)^{-\frac{D}{2}} \detZ e^{-\frac{1}{4T} x_{-} \Zz \cdot \cot \Zz \cdot x_{-}} \nonumber\\
 &&\times
 \Bigl\lbrack
 iT\,\textrm{tr}(\gbd \cdot \gbd^{\prime})  -\frac{1}{2}\xm \cdot \left(\cot \mathcal{Z} - \mathcal{Z}\cdot \csc^{2}\mathcal{Z}\right) \cdot \dot{\mathcal{G'}}_{B} \cdot \xm 
 + i x \cdot \dot{\mathcal{G}'}_{B} \cdot x^{\prime}
 \Bigr\rbrack \, .
\nonumber\\ 
\label{Sigmafin}
\ear
One would expect this to fulfill an equation generalizing the Gies-Karbstein equation \eqref{gieskarb}. And indeed, comparing \eqref{Sigmafin} with 
\eqref{L1scal} and the corresponding representation of the photonless scalar propagator in the field \cite{frgish-book,110}

\bear
D^{xx'}_{\rm scal} [F] &=& \int_{0}^{\infty}dT e^{-m^{2}T}(4\pi T)^{-\frac{D}{2}} \detZ e^{-\frac{1}{4T} x_{-} \Zz \cdot \cot \Zz \cdot x_{-}} \nonumber\\
\label{Dnophot}
\ear
it is easy to see that 

\begin{equation}
D_{\rm scal}^{xx'(1)1PR} = 	\frac{\partial D^{xx'}_{\rm scal}}{\partial F_{\mu\nu}}\frac{\partial \mathcal{L}^{(1)}_{\rm scal}}{\partial F^{\mu\nu}} 
+ \frac{ie}{2}D^{x x^{\prime}}_{\rm scal}x^{\mu}\frac{\partial \mathcal{L}^{(1)}_{\rm scal}}{\partial F^{\mu\nu}}x^{\prime \nu}  \, .
	\label{resComp}
\end{equation}
We remark that the form of the second term is specific for our choice of Fock-Schwinger gauge centered at $x'$. However, it is well-known how to convert our expression for the scalar propagator into an arbitrary gauge, as outlined in \cite{giesbook}.

\subsection{The propagator in momentum space}

Fourier transforming \eqref{resComp} we obtain the corresponding equation for the momentum space propagator:

\begin{equation}
D_{\rm scal}^{(1)1PR} = 	\frac{\partial D_{\rm scal}(p)}{\partial F_{\mu\nu}}\frac{\partial \mathcal{L}^{(1)}_{\rm scal}}{\partial F^{\mu\nu}}  \, .
	\label{resCompp}
\end{equation}
Here $D_{\rm scal}(p)$ denotes the constant field propagator in momentum space \cite{ditreuqed,110}

\bear
D_{\rm scal}(p)=
\int^{\infty}_{0}dT \, \e^{-m^2T}\, 
\e^{-T p \cdot \frac{{\rm tan} {\cal Z}}{\cal Z} \cdot p} 
\pdetZ  \, .
\ear
Note that the second term on the right-hand side of \eqref{resComp} does not survive the Fourier transform, since the momentum space
propagator depends only on a single momentum, so that the antisymmetric matrix $\frac{\partial \mathcal{L}^{(1)}_{\rm scal}}{\partial F^{\mu\nu}}$
cannot be saturated.  Explicitly, one finds

\bear
D_{\rm scal}^{(1)1PR}(p) &=&  \frac{ie^{2}}{2}\int_{0}^{\infty}dT^{\prime}(4\pi T^{\prime})^{-\frac{D}{2}} e^{-m^{2}T^{\prime}} \detZp \int_{0}^{\infty}dT T \, e^{-m^{2}T} \pdetZ e^{-T p \cdot \tZz \cdot p} \nonumber \\
	 && \times \Bigl\lbrack
		 T p\cdot  \frac{\sin {\cal Z}\cdot \cos {\cal Z} - {\cal Z}}{{\cal Z}^2 \cdot \cos^2 {\cal Z}}  \cdot \gbd^{\prime} \cdot  p +  \frac{1}{2} {\rm tr}( \tan {\cal Z} \cdot \dot{\cal G}'_B)
\Bigr\rbrack \, .
	 \label{Dpexpl}
\ear
We have also verified \eqref{Dpexpl} by a direct momentum space calculation, based on the momentum space 
worldline representation of the one-photon dressed scalar propagator given in \cite{110}. 

\section{Summary and outlook}

We have shown that a finite 1PR diagram, similar to the one found in \cite{giekar} 
for the two-loop EHL, exists for the scalar or electron propagator in a constant field already at the one-loop level.
For the scalar propagator, we have explicitly calculated this term in Fock-Schwinger gauge using the worldline
formalism, leading to equations generalizing the Gies-Karbstein equation \cite{giekar} in both $x$-space and
momentum space. The calculation for the spinor propagator is more involved and will be presented elsewhere. 

It will be interesting to see how this 1PR contribution affects quantities associated to the QED propagator in a constant
field such as the leading asymptotic $\ln^2 \frac{eB}{m^2}$ term in a strong magnetic field (see, e.g. \cite{kuzmik-book}
and refs. therein) or the famous Ritus mass shift \cite{ritusmass}. 
Clearly, the Gies-Karbstein mechanism will lead to a plethora of such terms at higher loop orders in constant-field QED.

Finally, we deem it important to stress the fact that the non-vanishing of the
Gies-Karbstein addendum relies on the photon being exactly massless. Therefore
the experimental verification of terms of this type might be promising as a way to improve on
the existing lower bounds on the photon mass.

\acknowledgments

We would like to thank Idrish Huet for helpful discussions. Both authors thank CONACYT for financial support through 
grant Ciencias Basicas 2014 No. 242461.


\end{document}